\documentclass[twocolumn,trackchanges]{aastex62}
\usepackage{amsmath}
\begin{document}

\title{Density-dependent van der Waals model under the GW170817 constraint}

% \correspondingauthor{August Muench}
% \email{greg.schwarz@aas.org, gus.muench@aas.org}

\author{O. Louren\c{c}o}
\affiliation{Departamento de F\'isica, Instituto Tecnol\'ogico de Aeron\'autica, DCTA, 
12228-900, S\~ao Jos\'e dos Campos, SP, Brazil}

\author{M. Dutra}
\affiliation{Departamento de F\'isica, Instituto Tecnol\'ogico de Aeron\'autica, DCTA, 
12228-900, S\~ao Jos\'e dos Campos, SP, Brazil}

\author{C. H. Lenzi}
\affiliation{Departamento de F\'isica, Instituto Tecnol\'ogico de Aeron\'autica, DCTA, 
12228-900, S\~ao Jos\'e dos Campos, SP, Brazil}

\author{M. Bhuyan}
\affiliation{Departamento de F\'isica, Instituto Tecnol\'ogico de Aeron\'autica, DCTA, 
12228-900, S\~ao Jos\'e dos Campos, SP, Brazil}
\affiliation{Department of Physics, Faculty of Science, University of Malaya, Kuala 
Lumpur 50603, Malaysia}
\affiliation{Institute of Research Development, Duy Tan University, Da Nang 550000, 
Vietnam.}

\author{S. K. Biswal}
\affiliation{Key Laboratory of Theoretical Physics, Institute of Theoretical Physics, 
Chinese 
Academy of Sciences, Beijing 100190, China.}

\author{B. M. Santos}
\affiliation{Universidade Federal do Acre, 69920-900, Rio Branco, AC, Brazil}

\begin{abstract}
We propose a density-dependent function for the attractive interaction in the original van 
der Waals model to correctly describe the flow constraint at the high-density regime of 
the symmetric nuclear matter. After a generalization to asymmetric nuclear matter, it was 
also possible to study the stellar matter regime from this new model. The mass-radius 
relation for neutron stars under $\beta$-equilibrium is found to agree with recent X-ray 
observations. The neutron star masses supported against gravity, obtained from some 
parametrizations of the model, are in the range of $(1.97-2.07)M_{\odot}$, compatible with 
observational data from the PSR J0348+0432 pulsar. Furthermore, we verify the reliability of the 
model in predicting tidal deformabilities of the binary system related to the GW170817 neutron star 
merger event and find a full agreement with the new bounds obtained by the LIGO/Virgo 
collaboration. 
\end{abstract}


\section{Introduction}

The applicability of hadronic equation(s) of state (EoS) goes from the description of 
superheavy nuclei to the structure of neutron stars. Therefore, a complete understanding 
of nuclear physics, as well as the astrophysics involved in these environments, is needed. 
The hadronic EoS is one of the most important ingredients to correctly predict, 
for instance, the mass of a neutron star. This object is one of the densest in the 
visible universe, having its density around $5-6$ times the nuclear saturation 
density~\citep{latt04}. It also provides a unique natural laboratory to test the hadronic 
EoS profile at the high-density regime~\citep{latt04}. Previously, some of the few 
constraints on the EoS, coming from the astrophysical context, were the values of 
the neutron star mass and the canonical star radius. Nowadays, the recent observation of 
the gravitational waves detected from a binary neutron star merger event, named as 
GW170817~\citep{ligonew} provides an opportunity to constrain various properties of 
hadronic EoS, and to search for more realistic ones, which will give a clear 
picture of the physics of a neutron star. The internal structure of such an object is a 
controversial subject since there are indications of many interesting phenomena like kaon 
production~\citep{glen98,glen99,gupt12,glen85}, hyperons 
emergence~\citep{glen98,glen91,amba60}, and the transition to a deconfinement quark 
phase~\citep{coll75}. Some of them are directly affected by the value of the bulk 
parameters, at the saturation density, given by the hadronic models, such as the symmetry 
energy ($J$), the incompressibility ($K_0$), the skewness parameter ($Q_0$), and related 
quantities. Many works establish constraints on these quantities, see, for instance, 
Ref.~\citep{rmf}. As an example, the incompressibility, which defines the stiffness of the 
EoS, has a value in the range of $K_0=(240\pm10$)~MeV, as found in 
Refs.~\citep{colo04,todd05,agra05}, predicted from the isoscalar giant monopole resonance 
of the heavy nuclei, or even the range of $250~\mbox{MeV}\leqslant K_0 \leqslant 
315~\mbox{MeV}$, more recently obtained in Ref.~\citep{stone} from a reanalysis 
of updated data on isoscalar giant monopole resonance energies of Sn and Cd isotopes. 
Some ranges for $K_0$ are found through a leptodermous expansion of the 
finite nucleus incompressibility, with $K_0$ as one of the terms. However, many works 
point out to the drawbacks of such a procedure, see, for instance, Refs.~\citep{k1,k2,k3}. 
The current consensus regarding the value of $K_0$ is $220~\mbox{MeV}\leqslant K_0 
\leqslant 260~\mbox{MeV}$, as one can see in a very recent review on this subject in 
Ref.~\citep{k4}, for instance. There are also a lot of uncertainties in the density 
dependence of some of the bulk parameters, such as the symmetry energy, especially at the 
high-density regime~\citep{se3,se1,se2,se4,se6,baldo,se5,se7}.

In this manuscript, we discuss the applicability of our proposed density-dependent van der 
Waals (\mbox{DD-vdW}) model, namely, an improved version of the previous van der Waals 
(vdW) model applied to nuclear matter~\citep{vov1,vov2,vov3}, that takes into account 
density-dependent repulsive and attractive interactions. Though the vdW model takes the 
description of the nuclear interactions in a simple way, its predictive capacity is as 
promising as other realistic relativistic~\citep{wale74,bogu77,sero79} and 
nonrelativistic models~\citep{skyr59,vaut72}, since it can reproduce some basic properties 
such as the saturation at a specific density, and the liquid-gas phase transition in 
symmetric nuclear matter. However, it presents some limitations like the causality 
violation at the low-density regime, which makes it inappropriate to describe stellar 
matter. In the \mbox{DD-vdW} model, this problem is circumvented. The neutron star 
calculations are performed, including the comparison of the tidal deformabilities of a 
binary neutron star system with the corresponding data related to the GW170817 event, 
recently reported in Ref.~\citep{ligonew}.  

This manuscript is organized as follows: in Sec.~\ref{den} we give an outline of the theoretical 
formalism that establishes the basis for a density-dependent version of the vdW model. In 
Sec.~\ref{app}, we discuss its causality limitation and show how its new version, proposed here, 
describes nuclear matter at higher densities, including the compatibility with the flow constraint 
predicted in Ref.~\citep{danielewicz}. A generalization to asymmetric systems is also developed, 
and special attention is given to stellar matter and tidal deformability calculations related to the 
GW170817 event. We finish the manuscript with a summary and concluding remarks in Sec.~\ref{con}.

\section{Density dependent vdW model}
\label{den}
By following the calculations performed at zero temperature regime in 
Refs.~\citep{vov1,vov2}, the energy density of an infinite symmetric nuclear matter (SNM) 
system in the grand canonical ensemble, can be obtained from the vdW model, as
\begin{eqnarray}
\mathcal{E}(\rho)=(1-b\rho)\mathcal{E}^*_{\rm id}(\rho^*) - a\rho^2,
\label{deev}
\end{eqnarray}
where $\mathcal{E}^*_{\rm id}$ is the kinetic energy of a relativistic ideal Fermi gas of 
nucleons 
of mass $M=938$ MeV, given by
\begin{eqnarray}
\mathcal{E}^*_{\rm id}(\rho^*) = 
\frac{\gamma}{2\pi^2}\int_0^{k_F^*}dk\,k^2(k^2+M^2)^{1/2},
\label{eidsym}
\end{eqnarray}
with $k_F^*=(6\pi^2\rho^*/\gamma)^{\frac{1}{3}}$, and $\rho^* = \rho/(1-b\rho)$. The 
degeneracy factor, $\gamma=4$ for SNM and $b = 16\pi r^3/3$, being $r$ the nucleon 
hard-sphere radius. The parameters $a$ and $b$ are related to the strength of attractive 
and repulsive interactions, respectively, and can be obtained by forcing the model to 
present a bound state at a particular density. In the case of infinite nuclear matter, 
the binding energy of such state is given by $\mathcal{E}(\rho_0)/\rho_0 - M = -B_0$ at 
the saturation density $\rho_0$. Usually, in the mean-field models, the binding energy 
per particle ($B_0 \approx 16.0$~MeV) at the saturation density $\rho_0 \approx 0.16$ 
fm$^{-3}$ are well established observables, and constrain the value of the parameters 
to \mbox{$a = 328.93$~MeV fm$^3$} and $b = 3.41$~fm${^3}$. 

From the structure of the excluded volume mechanism, it is clear that the vdW model has a 
 density range of $\rho < \rho_{\mbox{\tiny max}}$, with $\rho_{\mbox{\tiny max}}= 
b^{-1}$. For $b = 3.41$~fm${^3}$, one has $\rho_{\mbox{\tiny max}} = 1.83\rho_0$. In 
order to avoid such limitations, the repulsive term was modified in Ref.~\citep{vov3}, by 
applying the Carnahan-Starling~(CS)~\citep{cs} method of excluded volume, in which the 
pressure of hard-core particles of radius $r$ is given by $P=\rho T Z_{\rm CS} (\eta)$, 
with  
\begin{eqnarray} 
Z_{\rm CS} (\eta) = \frac{1 + \eta + \eta^2 - \eta^3}{(1-\eta)^3} 
= 1 + \sum_{j=0}^\infty(j^2+3j)\eta^j,
\end{eqnarray}
and $\eta=b\rho/4$. By following this method, the first eight virial expansion coefficients are 
obtained, unlike the traditional vdW excluded volume procedure, where only two of them are recovered 
since for this case $Z(\eta)=(1-4\eta)^{-1}$. By taking the CS procedure to the vdW model, the 
nuclear matter energy density in the grand canonical ensemble is written 
as~\citep{vov3}\begin{eqnarray}
\mathcal{E}(\rho)=f(\eta)\mathcal{E}^*_{\rm id} - a\rho^2,
\label{decs}
\end{eqnarray}
with $f(\eta)= e^{-(4-3\eta)\eta/(1-\eta)^2}$ and $\rho^*=\rho/f(\eta)$. We name this 
model as \mbox{vdW-CS} one. For this model, it is found \mbox{$a=347.02$~MeV fm$^3$} and 
$b=4.43$~fm${^3}$ with maximum density given by $\rho_{\mbox{\tiny max}} = 
4b^{-1}=5.64\rho_0$~\citep{vov3}.

In an effective way, the \mbox{vdW-CS} model can be seen as a density-dependent model by 
rewriting Eq.~(\ref{decs}) as  
\begin{eqnarray}
\mathcal{E}(\rho)=[1-b(\rho)\rho]\mathcal{E}^*_{\rm id}(\rho^*) - a(\rho)\rho^2,
\label{dedd}
\end{eqnarray}
with $\rho^* = \rho/[1-b(\rho)\rho]$ and 
\begin{eqnarray}
b(\rho)=\frac{1-f(\eta)}{\rho} = \frac{1}{\rho}-\frac{1}{\rho}{\rm exp}
\left[\dfrac{-(4-\frac{3b\rho}{4})\frac{3b\rho}{4}}{\left(1-\frac{3b\rho}{4}\right)^2}
\right].
\label{brho}
\end{eqnarray}
Thus, the repulsive interaction becomes a density-dependent function. We generalize this 
idea to the attractive part, regulated by the $a$ parameter, by making it depending on 
$\rho$ as well, i.e., we assume $a\rightarrow a(\rho)$ in Eq.~(\ref{dedd}). Therefore, 
the original vdW model is given by the particular case in which $b (\rho) = b$ and $a 
(\rho) = a$. For the \mbox{vdW-CS} model, $b(\rho)$ is given by Eq.~(\ref{brho}) and $a 
(\rho) = a$. 

From the perspective of a density-dependent model, it is also possible to use 
Eq.~(\ref{dedd}) to describe all real gases models analyzed in Ref.~\citep{vov3}, namely, 
Redlich-Kwong-Soave~(RKS), Peng-Robinson~(PR), and Clausius-2~(C2), by identifying
\begin{eqnarray}
a(\rho)&=&\frac{a}{b\rho}{\rm ln}(1 + b\rho)\hspace{3cm} {\rm (RKS)},
\label{arhorks} \\
a(\rho)&=&\frac{a}{2\sqrt{2}b\rho}{\rm 
ln}\left[\frac{1+b\rho(1+\sqrt{2})}{1+b\rho(1-\sqrt{2})}\right]\hspace{0.65cm}
{\rm (PR)},
\label{arhopr}
\end{eqnarray}
and
\begin{eqnarray}
a(\rho)&=&\frac{a}{1+b\rho}\hspace{3.9cm} {\rm (C2)}.
\label{arhoc2}
\end{eqnarray}
We denote here the last model by Clausius-2 (C2) because it is a two parameters ($a$, 
$b$) version of the original Clausius model. A three parameters ($a$, $b$, $c$) version of 
such model is studied in Ref.~\citep{vov4}. 

From Eq.~(\ref{dedd}), the pressure of the system can be written as
\begin{eqnarray}
P(\rho) &=& \rho^2\frac{\partial(\mathcal{E}/\rho)}{\partial\rho}
= P^*_{\rm id} - a(\rho)\rho^2 + \rho\Sigma(\rho),
\end{eqnarray}
with
\begin{eqnarray}
P^*_{\rm id} = \frac{\gamma}{6\pi^2}\int_0^{k_F^*}\frac{dk\,k^4}{(k^2+M^2)^{1/2}}.
\label{pidsym}
\end{eqnarray}
Here, $a'$ and $b'$ are the first density derivatives of $a (\rho)$ and $b (\rho)$,  
respectively. Furthermore, the density dependence of the interactions gives rise to a 
rearrangement term in the pressure given by $\Sigma(\rho) = b'\rho P^*_{\rm id} - 
a'\rho^2$. The incompressibility of SNM is calculated as the derivative of pressure with 
respect to $\rho$ as 
\begin{eqnarray}
&K&(\rho)=9\frac{\partial P}{\partial\rho}\nonumber
\\
&=& \frac{1+b'\rho^2}{[1-b(\rho)\rho]^2}K^*_{\rm id} 
- 9\rho[2a(\rho) + a'\rho] + 9[\Sigma(\rho) + \rho\Sigma'],\nonumber\\
\end{eqnarray}
where $K^*_{\rm id} = 3k_F^{*2}/\sqrt{k_F^{*2}+M^2}$, and
\begin{eqnarray}
\Sigma'(\rho) &=& (b''\rho + b')P^*_{\rm id} + 
\frac{(1+b'\rho^2)b'\rho}{9[1-b(\rho)\rho]^2}K^*_{\rm id} \nonumber\\
&-& a''\rho^2 - 2a'\rho.
\end{eqnarray}
Finally, the chemical potential is obtained from
\begin{align}
\mu(\rho) &= \frac{\partial\mathcal{E}}{\partial\rho} = \mu^*_{\rm id} + b(\rho)P^*_{\rm 
id} + 
\Sigma(\rho) - 2a(\rho)\rho,
\end{align}
with $\mu^*_{\rm id}=E_F^*=(k_F^{*2}+M^2)^{1/2}$. We remind the readers that the  
rearrangement term $\Sigma (\rho)$, originated from the density dependence of the 
attractive and repulsive interactions, is an essential quantity in order to preserve the 
thermodynamic consistency of the model. It is straightforward to see that equations of 
state presented here with $\Sigma (\rho)$ included, give rise to the relation $P + 
\mathcal{E} = \mu\rho$, indicating the consistency.

\section{Results and Discussions}
\label{app}

Since the structure and EoS of the density-dependent vdW model were presented, we are now able to 
analyze in more details the real gases of Ref. \citep{vov3}, namely, those given by 
Eqs.~(\ref{arhorks})-(\ref{arhoc2}). At present, a diverse set of nonrelativistic and relativistic 
mean-field (RMF) models describe quite well nuclear and neutron star 
matter~\citep{rmf,skyrme,ijmpe1,ijmpe2}. Hence, it is also important to see how the 
density-dependent vdW models describe the same environment.

\subsection{Causality analysis from flow constraint}
Our first analysis is concerning the flow constraint proposed in 
Ref.~\citep{danielewicz}, in which limits on the pressure-density relationship of SNM and its 
curvature can be obtained from the experimental data on the motion of ejected matter in the 
energetic nucleus-nucleus collisions. Measurements of the particle flow in the collisions of 
\mbox{$^{197}\rm Au$} nucleus at incident kinetic energy per nucleon varying from about $0.15$ to 
$10$ GeV is used in Ref.~\citep{danielewicz}, that extrapolated for pressure at the range of $2.0 
\leqslant \rho/\rho_0 \leqslant 4.6$, at the zero temperature. In principle, the CS excluded volume 
procedure enables the models to reach values for $\rho_{\mbox{\tiny max}}$ greater than those found 
by using the traditional excluded volume method. However, the causality is violated at densities 
lower than $\rho_{\mbox{\tiny max}}$.

We remind the reader that the conventional excluded volume technique for nucleons are 
taken into account for repulsion at short distances by treating them as rigid spheres in a 
nonrelativistic context. In a relativistic framework, as in the case of nuclear matter 
at high densities, the Lorentz contraction of such hard sphere nucleons should be 
implemented to avoid causality violation for any density regions~\citep{causality}. Thus, 
the effect of Lorentz contraction in nucleons can be seen as a decreasing function of 
their radius. Effectively, this is the case of the CS method, since it treats the excluded 
volume parameter as a density-dependent quantity, 
see Eq.~(\ref{brho}). The method identifies the nucleon as a 
sphere in which the radius is depending on the density, at least as a first approach. 
However, such a density-dependent is not enough to completely avoid the causality 
violation. The sound velocity is still greater than $1$ (units of $c=1$) in CS approach, 
but at higher densities in comparison with the case in which conventional excluded volume 
procedure (fixed radius) is taken.  

\begin{figure}[!htb]
\centering
\includegraphics[scale=0.33]{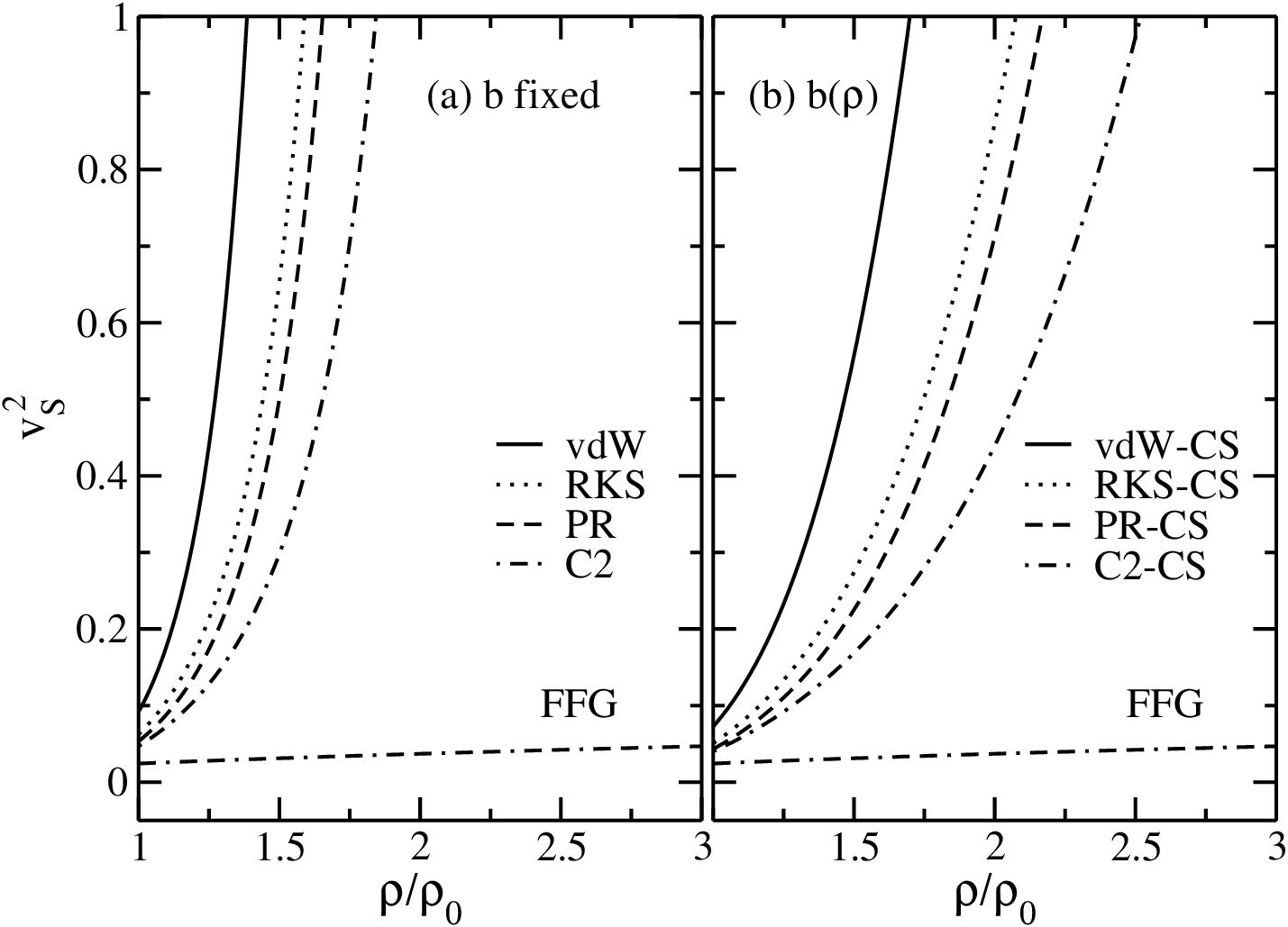}
\vspace{-0.2cm}
\caption{$v^2_s$ as a function of $\rho/\rho_0$ for real gases for SNM at zero temperature 
for (a) fixed excluded volume parameter, and (b) Carnahan-Starling method. Lower curve: 
free Fermi gas of 
massive nucleons.}
\label{vs2}
\end{figure}

In Fig.~\ref{vs2}, we show the squared sound velocity, $v_s^2=\partial P/\partial\mathcal{E} = 
K/9\mu$, as a function of the ratio $\rho/\rho_0$ for the real gases models presented in the last 
section.  

It is clear that the repulsive interaction plays an important role in 
the causal limit, since it induces its violation at higher densities. The physical reason 
of such a result is that the CS method weakens the repulsive interaction as a function 
of density, producing results closer to the ones of an ideal gas of massive point 
nucleons. The more $b(\rho)$ decreases, the more the nucleons behave like point 
particles, since $b(\rho)\rightarrow 0$ indicates structureless objects. 

\begin{figure}[!htb]
\centering
\includegraphics[scale=0.33]{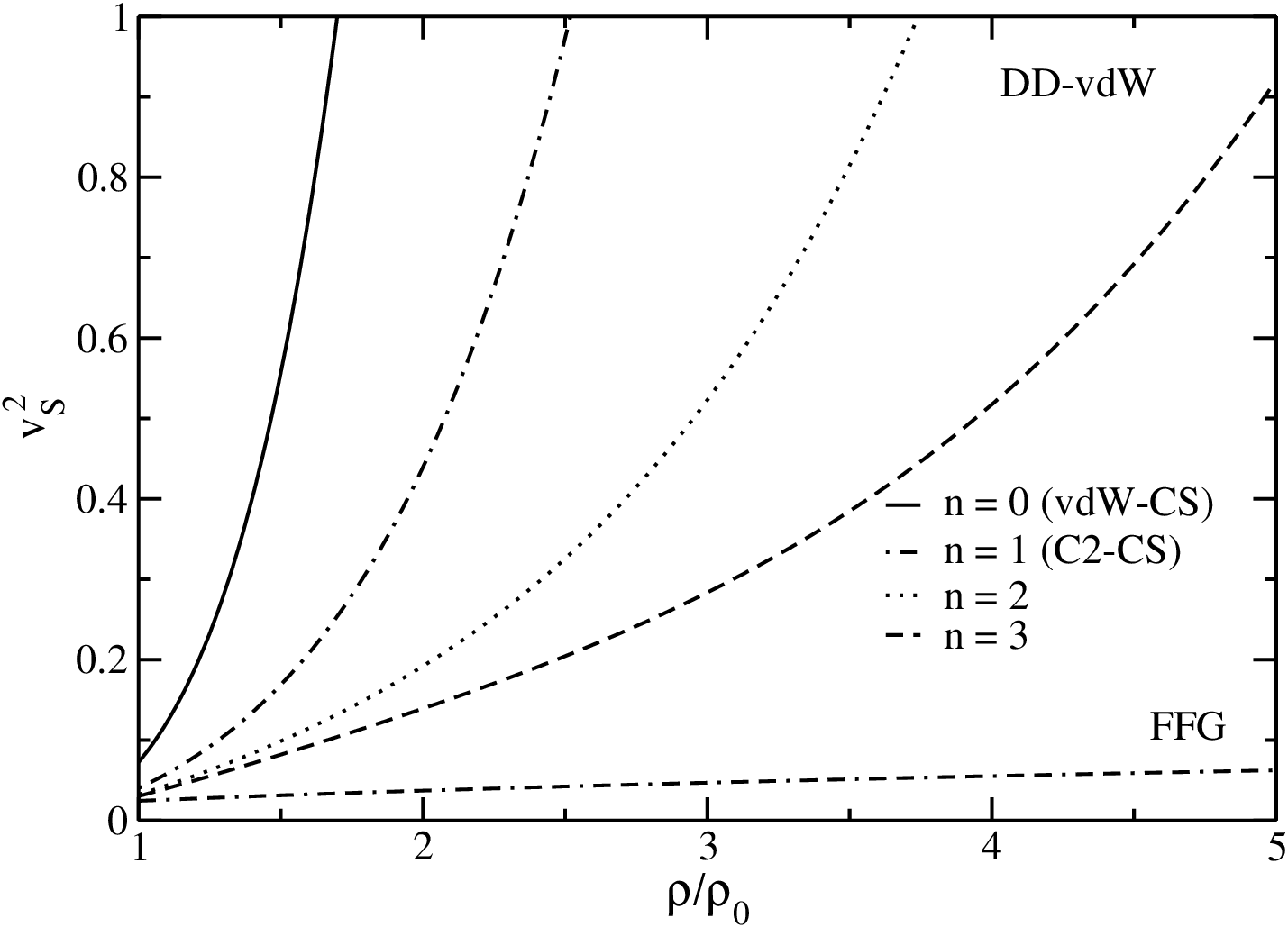}
\vspace{-0.2cm}
\caption{Squared sound velocity as a function of $\rho/\rho_0$ for the \mbox{DD-vdW} 
model. Results for SNM at zero temperature. Lower curve: free Fermi gas of massive 
nucleons.}
\label{vs2dd}
\end{figure}

Even by applying the CS excluded volume method in the real gases models, the causality is 
still broken in the range of $2.0\leqslant\rho/\rho_0\leqslant 4.6$ of the flow 
constraint. In Fig.~\ref{vs2} we see that $v^2_s=1$ at $\rho/\rho_0 = 2.5$ for the
\mbox{C2-CS} model, for instance. In order to circumvent this limitation, we implement a 
modification in the attractive interaction of the model. A new proposed form for the 
parameter $a(\rho)$, is given by,
\begin{eqnarray}
a(\rho)&=&\frac{a}{(1+b\rho)^n}.
\label{arhodd}
\end{eqnarray}
It is inspired in the \mbox{C2-CS} model, in which causality is violated at higher densities in 
comparison with the remaining models. Its repulsive interaction remains the same, i.e., 
the CS excluded volume method is considered. The model with the new proposal for 
$a(\rho)$ is named here as the \mbox{DD-vdW} model, with the couplings $a(\rho)$ and 
$b(\rho)$ given by Eqs.~(\ref{arhodd}) and (\ref{brho}), respectively. Notice that 
for the particular cases of $n=0$, and $n=1$, the \mbox{vdW-CS} and \mbox{C2-CS} models 
are reproduced, respectively. The squared sound velocity for some values of $n$ 
is displayed in Fig.~\ref{vs2dd}. 

Notice that the effect of the $n$ power in $a(\rho)$, Eq.~(\ref{arhodd}), is to weaken the 
strength of the attractive interaction. The more $a(\rho)$ decreases, the more the 
\mbox{DD-vdW} model approaches to the free Fermi gas of massive particles. The combined 
effects of the density-dependent parameters $a(\rho)$ and $b(\rho)$ favor the model to 
move the causality violation to higher densities, as one can see in Fig.~\ref{vs2dd}.

\subsection{DD-vdW model in symmetric nuclear matter}

\begin{figure}[!htb]
\centering
\includegraphics[scale=0.33]{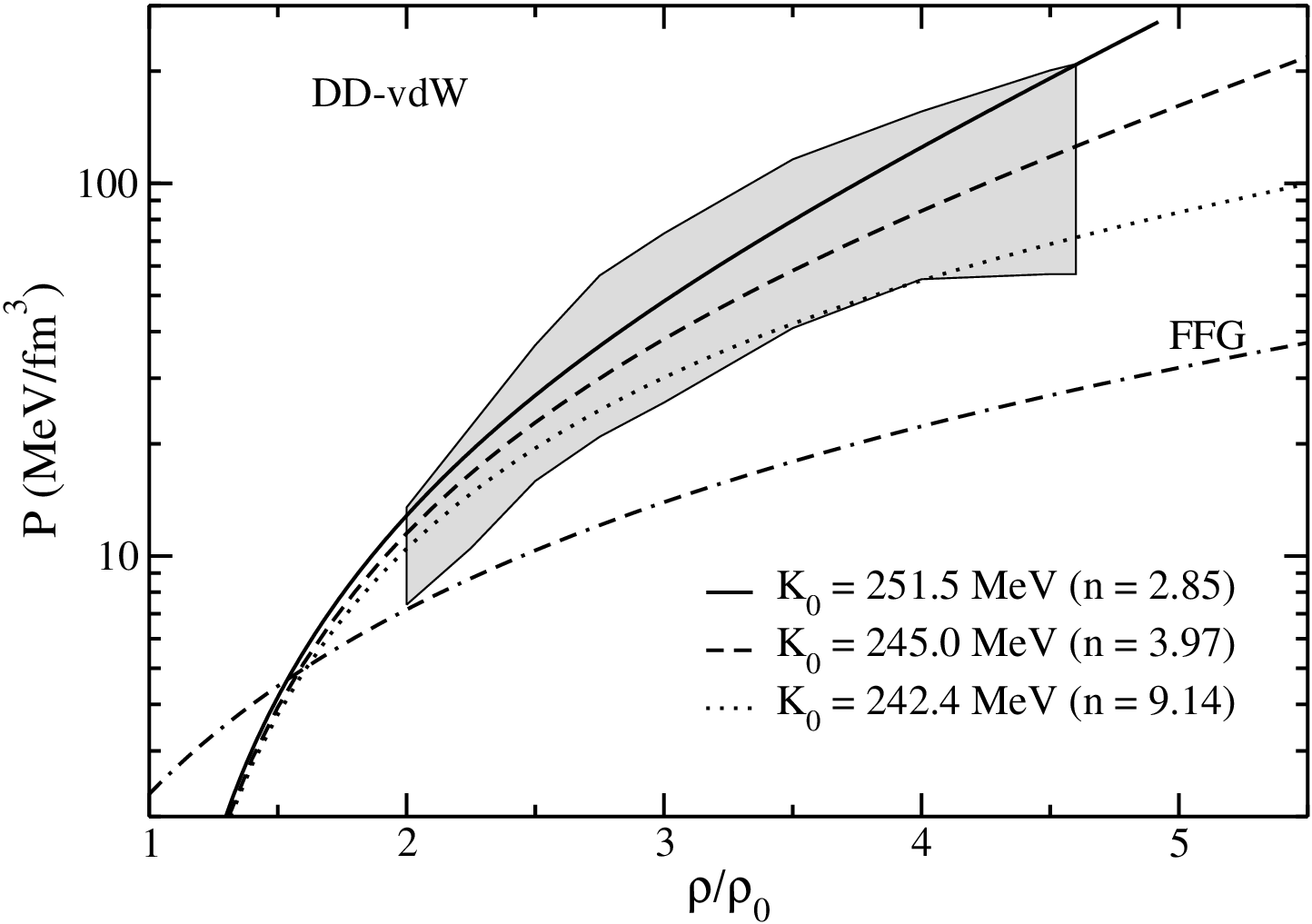}
\vspace{-0.2cm}
\caption{Pressure as a function of $\rho/\rho_0$ for different \mbox{DD-vdW} 
parametrizations (different $K_0$ values). Bands: flow constraint described in 
Refs.~\citep{rmf,danielewicz}. Lower curve: free Fermi gas of massive 
nucleons.}
\label{flow}
\end{figure}

The $n$ power in Eq.~(\ref{arhodd}) directly affects the incompressibility value at the 
saturation density, $K_0\equiv K(\rho_0)$. Therefore, we can use this parameter to control 
the $K(\rho_0)$ quantity. We use this procedure to submit the \mbox{DD-vdW} model to the 
flow constraint of Ref.~\citep{danielewicz}. The resulting parametrizations of this model 
with different $K_0$ values are shown in Fig.~\ref{flow}. 

From this figure, we see that the \mbox{DD-vdW} model satisfies the flow constraint for 
parametrizations presenting $242.4~\mbox{MeV}\leqslant K_0 \leqslant 251.5~\mbox{MeV}$.  All the 
curves of Fig.~\ref{flow} are compatible with the causal limit, i.e., for each curve exhibited one 
has $v_s^2<1$. The parametrization with $K_0=251.5$ MeV, for instance, is causal for 
$\rho/\rho_0\leqslant 4.9$. Furthermore, all these parametrizations are also in agreement with the 
restriction of $220~\mbox{MeV}\leqslant K_0 \leqslant 260~\mbox{MeV}$~\citep{k4}. It is worth to 
mention that in the recent work of Ref.~\citep{vov3}, the results pointed out that none of the 
other real gases models produces $K_0$ inside the aforementioned range for $K_0$. Finally, it is 
also clear from Fig.~\ref{flow} that the weakening effect of the interactions leads the 
\mbox{DD-vdW} model to the direction of the free Fermi gas, with 
full agreement with the flow constraint. Another approach performed in the van der Waals model that 
also makes it to satisfy the flow constraint, and also the maximum mass observational data for 
neutron stars, includes induced surface tension in its formulation. For details, see 
\citep{Sagun:2017eye,Bugaev:2018uum} and references therein.

\subsection{Asymmetric matter formulation}

In order to proceed for a complete analysis of the infinite nucleonic matter, it is 
necessary to take the isospin asymmetry system into account, i.e., the EoS for 
$y\equiv\frac{\rho_p}{\rho}\neq \frac{1}{2}$. Here $\rho_p$ is the proton density. For the 
original vdW model, a generalization for different hard sphere particles was performed 
in Ref.~\citep{multi}, where EoS were developed for neutron-proton and 
\mbox{nucleons-$\alpha$} particles systems. Here, in order to avoid the emergence of more 
free parameters, we consider the same density-dependent couplings $a(\rho)$ and $b(\rho)$ 
of the symmetric matter, Eqs.~(\ref{arhodd}) and (\ref{brho}), for protons and neutrons. 
Therefore, the individual components are distinguished only by their respective kinetic 
energies. However, as already discussed in Ref.~\citep{multi} in the case of $a(\rho)=a$ 
and $b(\rho)=b$, the symmetry energy at the saturation density calculated from this 
approach, $J\equiv \mathcal{S}(\rho_0)$, presents an underestimate value. In order to 
avoid such a limitation, we propose a new term in the energy density of the 
\mbox{DD-vdW} model, proportional to the squared difference between protons and neutrons 
densities, $\rho_3 = \rho_p -\rho_n=(2y-1)\rho$, as widely used in some RMF 
models~\citep{rmf}. This new term can be seen as a simulation of the $\rho$ meson exchange 
in its simple form, i.e., a minimal coupling between this meson and the finite particle 
nucleon (excluded volume included). Thus, the asymmetric nuclear matter energy density is 
given by,
\begin{align}
\mathcal{E}(\rho,y)&=[1-b(\rho)\rho]\mathcal{E}^*_{\rm id}(\rho^*_p,\rho^*_n) - 
a(\rho)\rho^2 + d\rho_3^2, 
\label{deddy}
\end{align}
where $\mathcal{E}^*_{\rm id}(\rho^*_p,\rho^*_n) = \mathcal{E}^{*p}_{\rm id}(\rho^*_p) 
+ \mathcal{E}^{*n}_{\rm id}(\rho^*_n)$, for $\mathcal{E}^{*i}_{\rm id}(\rho^*_i)$ 
following the same form as in Eq.~(\ref{eidsym}) with $\gamma=2$, $k_F^*\rightarrow 
k_F^{*i}$ and $\rho^*\rightarrow \rho^*_i$ ($i=p,n$).  The different densities are 
related 
to each other by
\begin{eqnarray} 
\rho^*_p = \frac{\rho_p}{1-b(\rho)\rho},\qquad \rho^*_n = \frac{\rho_n}{1-b(\rho)\rho}. 
\end{eqnarray}
The couplings are written as in the previous case, see Eqs.~(\ref{brho}) and 
(\ref{arhodd}), respectively, for $b(\rho)$ and $a(\rho)$. 

From Eq. (\ref{deddy}), we can derive  the remaining quantities for the asymmetric 
nuclear matter. The expressions are,
\begin{eqnarray}
P(\rho,y) = P^*_{\rm id} - a(\rho)\rho^2 + \rho\Sigma(\rho,y) 
+ d(2y - 1)^2\rho^2,\quad
\label{pddy}
\end{eqnarray}
and
\begin{align}
\mu_{p,n}(\rho,y) = \frac{\partial\mathcal{E}}{\partial\rho_{p,n}} 
&= \mu^{*p,n}_{\rm id} + b(\rho)P^*_{\rm id}(\rho^*_p,\rho^*_n) + \Sigma(\rho,y) 
\nonumber\\
&- 2a(\rho)\rho \pm 2d(2y - 1)\rho,
\label{mupdd}
\end{align}
for the pressure, and chemical potentials for protons (upper sign) and neutrons (lower 
sign), respectively. Furthermore, $P^*_{\rm id}(\rho^*_p,\rho^*_n) = P^{*p}_{\rm 
id}(\rho^*_p) + P^{*n}_{\rm id}(\rho^*_n)$ is written as in Eq.~(\ref{pidsym}) with 
$\gamma=2$, $k_F^*\rightarrow k_F^{*i}$ and $\rho^*\rightarrow \rho^*_i$. The ``ideal'' 
chemical potentials are $\mu^{*i}_{\rm id} = E^{*i}_F$. The rearrangement term and its 
derivative with respect to the density are given by $\Sigma(\rho,y) = b'\rho P^*_{\rm 
id}(\rho^*_p,\rho^*_n) - a'\rho^2$, and
\begin{align}
\Sigma'(\rho,y) &= \frac{(1+b'\rho^2)b'\rho}{9[1-b(\rho)\rho]^2}[yK^{*p}_{\rm 
id}(\rho^*_p) + (1-y)K^{*n}_{\rm id}(\rho^*_n)] \nonumber\\
&+ (b''\rho + b')P^*_{\rm id}(\rho^*_p,\rho^*_n) - a''\rho^2 - 2a'\rho,
\end{align}
with $K^{*i}_{\rm id}(\rho^*_i)$ defined as in the symmetric nuclear matter case.

By defining the \mbox{DD-vdW} model generalized to asymmetric matter in this way, there 
are only four free parameters to be adjusted, namely, $a$, $b$, $n$, and $d$. The first 
three ones are already determined from the symmetric case (reproducing the values 
of $\rho_0$, $B_0$, and $K_0$). The remaining free parameter, $d$, is adjusted in order to 
correctly reproduce a bulk quantity of asymmetric nuclear matter, namely, the symmetry 
energy. This quantity measures the change in binding of the nucleon system as the proton 
to neutron ratio is changed at a fixed value of the density, $\mathcal{S}(\rho) = 
E(\rho,0) - E(\rho,1/2)$, where $E(\rho,y)$ is the energy per particle. A detailed 
analysis of the quantity is quite important for understanding many aspects of different 
isospin asymmetric systems, from astrophysics to finite nuclei. Furthermore, the symmetry 
energy slope at saturation density provides the dominant contribution to the pressure in 
neutron stars, as well as affects the neutron skin thicknesses of heavy nuclei 
\citep{pieka,bhu15,bhu18}. For a recent review regarding the importance of $\mathcal{S} 
(\rho)$, see Ref.~\citep{baldo}. By considering $E(\rho,y)\simeq E(\rho,1/2) + 
\mathcal{S}_2(\rho)(1-2y)^2 + \mathcal{O}[(1-2y)^4]$, one can take 
$\mathcal{S}(\rho)\simeq \mathcal{S}_2(\rho)$ as a good approximation in order to compute  
the symmetry energy. The $\mathcal{S} (\rho)$ of the \mbox{DD-vdW} model can be written 
as, 
\begin{align}
\mathcal{S}(\rho) \simeq\mathcal{S}_2(\rho) =
\frac{1}{8}\frac{\partial^{2}(\mathcal{E}/\rho)}{\partial y^{2}}\Big|_{y=\frac{1}{2}} 
= \mathcal{S}_{kin}^*(\rho) + d\rho,
\label{esym}
\end{align}
with $\mathcal{S}_{kin}^*(\rho)=k_F^{*2}/(6E_F^*)$ and $k_F^* =  
(3\pi^2\rho^*/2)^{\frac{1}{3}}$. 

One can notice that the excluded volume procedure directly affects the kinetic part of 
$\mathcal{S}(\rho)$ as well as all in other thermodynamical quantities. The determination 
of the $d$ parameter is straightforward since we use the analytical expression of 
$\mathcal{S}(\rho)$ for this aim by imposing our model to present consistent values for 
the symmetry energy at saturation density, $J$. We find $d$ by constraining $J$ to the 
range of $25~\mbox{MeV}\leqslant J \leqslant 35~\mbox{MeV}$. This range was 
established in order to encompass data obtained from various terrestrial nuclear 
experiments and astrophysical observations, such as isospin diffusion, neutron skins, 
pygmy dipole resonances, modes of decay near the drip-line, transverse flow, mass-radius 
relations, and torsional crust oscillations of neutron stars. One can find a collection 
of these data in Ref.~\citep{jl}. 

The slope parameter, i.e., the symmetry energy slope as a function of density, is 
obtained from Eq.~(\ref{esym}) as 
\begin{equation}
L(\rho) = 3\rho\frac{\partial\mathcal{S}}{\partial\rho} 
= \xi(\rho)L_{kin}^*(\rho) + 3d\rho,
\end{equation}
where
\begin{eqnarray}
L_{kin}^*(\rho)=\frac{k_F^{*2}}{3E^*_F}\left(1 - \frac{k_F^{*2}}{2E_F^{*2}}\right) 
= 2\mathcal{S}_{kin}^*\left(1 - \frac{3\mathcal{S}_{kin}^*}{E_F^*}\right)\quad
\end{eqnarray}
and $\xi(\rho) = [{1 + b'\rho^2}]/[{1 - b(\rho)\rho}]$. The advantage of the specific form 
of the last term included in the Eq.~(\ref{deddy}) is the possibility of an analytical 
relationship between $\mathcal{S}(\rho)$ and $L(\rho)$ for all densities. Notice that 
because $d=(\mathcal{S}-\mathcal{S}_{kin}^*)/\rho$, it is possible to write
\begin{eqnarray}
L(\rho) = 3\mathcal{S}(\rho) + \mathcal{S}_{kin}^*(\rho)\left\lbrace
2\xi\left[ 1 - \frac{3\mathcal{S}_{kin}^*(\rho)}{E_F^*(\rho)}\right] - 3 
\right\rbrace.\quad
\label{clgr}
\end{eqnarray}
At the saturation density, this equation is used to calculate $L(\rho_0) \equiv L_0$. The quantity 
$L_0$ is of great interest for constraining the EoS of asymmetric nuclear matter in many hadronic 
models \citep{baldo,bianca2,bianca,lim}. For the \mbox{DD-vdW} model, we find $L_0$ in the range of 
$63.4~\mbox{MeV}\leqslant L_0 \leqslant 96.5~\mbox{MeV}$, by taking into account the constraint of 
$25~\mbox{MeV}\leqslant J \leqslant 35~\mbox{MeV}$ and the range of $242.4~\mbox{MeV}\leqslant K_0 
\leqslant 251.5~\mbox{MeV}$ obtained from the flow constraint analysis. The obtained values for 
$L_0$ are in agreement with the constraint of $25~\mbox{MeV}\leqslant L_0 \leqslant 115~\mbox{MeV}$ 
used in Refs.~\citep{rmf,jl}.

\subsection{Stellar matter}

A neutron star (NS) is a very compact object composed not only by neutrons, but also by 
protons and leptons. Different reactions such as the $\beta$ decay, namely, $n\rightarrow 
p + e^- + \bar{\nu}_e$ and its inverse process $p+ e^-\rightarrow n + \nu_e$, take place 
in the interior of a NS. For densities in which the electron chemical potential exceeds 
the muon mass value, the reactions $e^-\rightarrow \mu^- + \nu_e + \bar{\nu}_\mu$, $p + 
\mu^-\rightarrow n + \nu_\mu$ and $n\rightarrow p + \mu^- + \bar{\nu}_\mu$ may be 
energetically allowed. In this case, muons can also emerge. Here, we consider that all 
neutrinos escape from the star. By taking these assumptions into account, one can write 
total energy density and pressure of the stellar system, respectively, as
\begin{eqnarray}
\mathcal{E}_T(\rho,\rho_e,y) &=& \mathcal{E}(\rho,y) + \frac{\mu_e^4(\rho_e)}{4\pi^2}
\nonumber\\
&+& 
\frac{1}{\pi^2}\int_0^{\sqrt{\mu_\mu^2(\rho_e)-m^2_\mu}}dk\,k^2(k^2+m_\mu^2)^{1/2},\qquad
\label{det}
\end{eqnarray}
and
\begin{eqnarray} 
P_T(\rho,\rho_e,y) &=& P(\rho,y) + \frac{\mu_e^4(\rho_e)}{12\pi^2}
\nonumber\\
&+& \frac{1}{3\pi^2}\int_0^{\sqrt{\mu_\mu^2(\rho_e)-m^2_\mu}} 
\frac{dk\,k^4}{(k^2+m_\mu^2)^{1/2 } },\qquad
\label{presst}
\end{eqnarray}
where, $\mathcal{E}(\rho,y)$ and $P(\rho,y)$ are given in the Eqs.~(\ref{deddy}) and  
(\ref{pddy}), respectively. The chemical equilibrium and the charge neutrality conditions 
are given by $\mu_n(\rho,y) - \mu_p(\rho,y) = \mu_e(\rho_e)$ and $\rho_p(\rho,y) - \rho_e 
 = \rho_\mu(\rho_e)$, with $\mu_p$ and $\mu_n$ defined in Eq.~(\ref{mupdd}). Furthermore, 
one has $\mu_e=(3\pi^2\rho_e)^{1/3}$, $\rho_p=y\rho$, \mbox{$\rho_\mu=[(\mu_\mu^2 - 
m_\mu^2)^{3/2}]/(3\pi^2)$}, and  $\mu_\mu=\mu_e$, for $m_\mu=105.7$~MeV and massless 
electrons. Thus, for each input density $\rho$, the quantities $\rho_e$ and $y$ are 
calculated by solving the restrictions for the chemical potentials and densities 
simultaneously. The output is used to compute $\mathcal{E}_T(\rho,\rho_e,y)$ and 
$P_T(\rho,\rho_e,y)$ as a function of the density. 

\begin{figure}[!htb]
\centering
\includegraphics[scale=0.33]{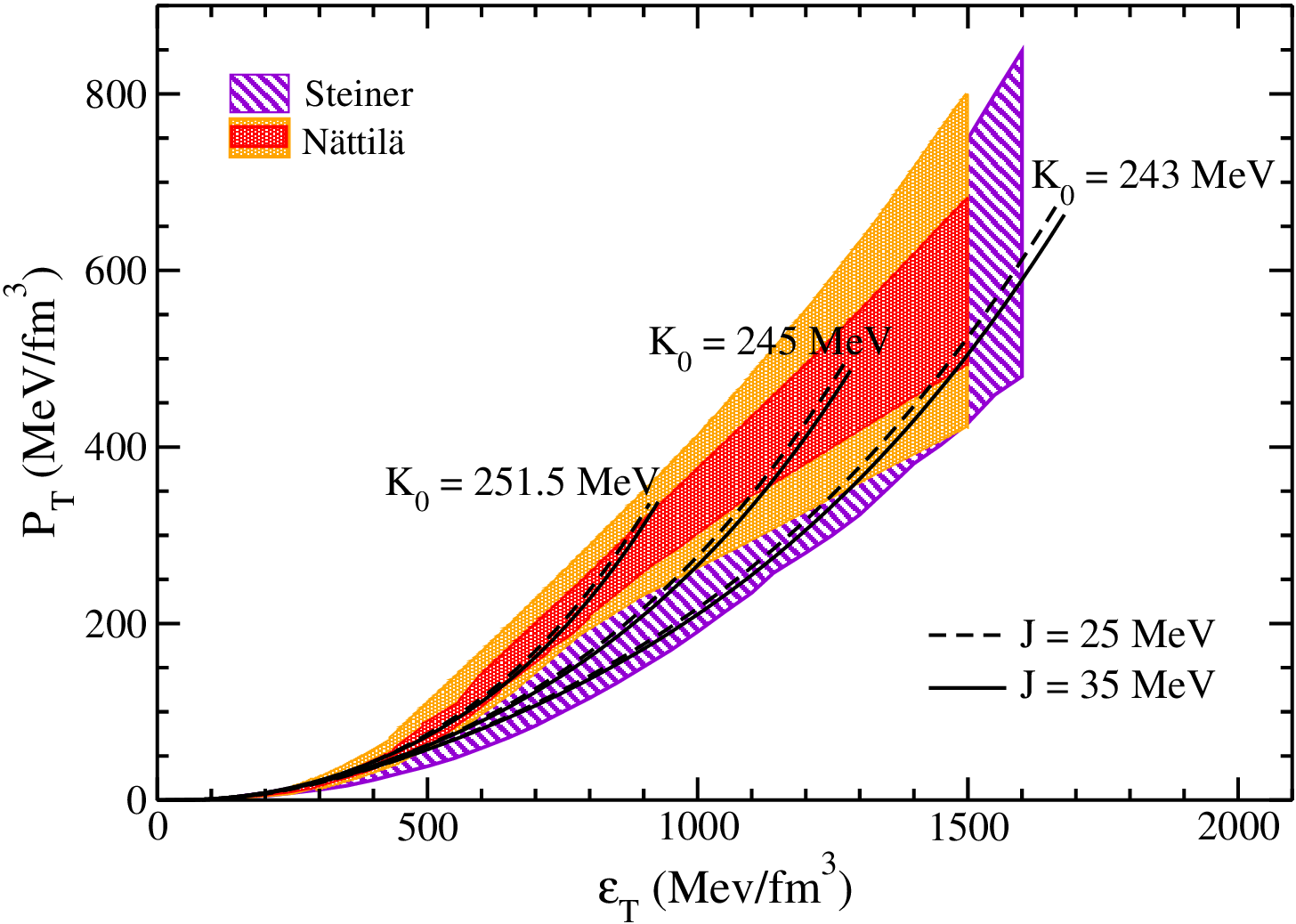}
\vspace{-0.2cm}
\caption{Total pressure versus total energy density for the \mbox{DD-vdW} 
parametrizations in $\beta$ equilibrated matter. Violet band region: calculations 
extracted from the Ref.~\citep{q01}. Red and orange one: limits found in 
Ref.~\citep{nat}.}
\label{pe}
\end{figure}

In Fig.~\ref{pe}, we show the EoS of neutron star matter under $\beta$-equilibrium  for 
\mbox{DD-vdW} parametrizations consistent with the flow constraint. All the 
$P_T\times \mathcal{E}_T$ curves in the figure are consistent with the causal limit, i.e., 
the parametrizations are restricted to density ranges in which $v_s^2<1$. In these curves, 
one can observe the effects of the bulk parameters. For example, it is clear that $K_0$ 
plays a major role in the EoS of the neutron star matter, exactly as in the flow 
constraint, see Fig.~\ref{flow}. A more careful inspection shows that the results 
marginally depend on the symmetry energy. It is also verified that the curves are also 
compatible with the results found by Steiner~\citep{q01} and N\"attili\"a \citep{nat}. 

\begin{figure}[!htb]
\centering
\includegraphics[scale=0.33]{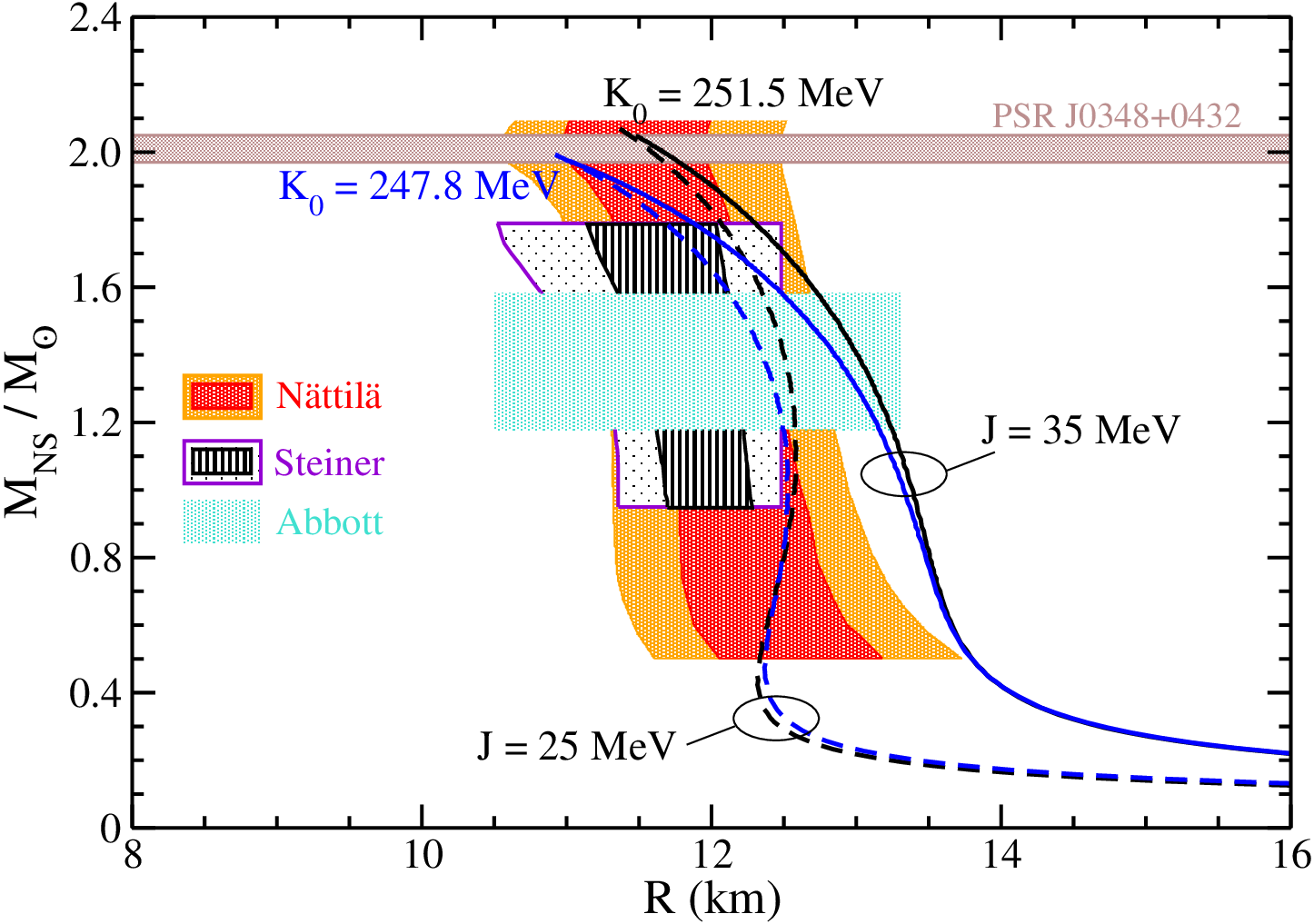}
\vspace{-0.2cm}
\caption{Mass-radius diagrams for some \mbox{DD-vdW} parametrizations. The horizontal 
brown band indicates the masses of the PSR~J038+0432~\citep{psr2} pulsar. Outer orange and 
inner red bands: data extracted from Ref.~\citep{nat}. Outer white and inner black bands: 
data extracted from Ref.~\citep{q01}. Turquoise band: limits from the GW170817 event found 
in Ref.~\citep{ligonew}. $M_\odot$ is the solar mass.}
\label{mr}
\end{figure}

The mass-radius diagrams of spherically symmetric neutron stars are found from the solutions of the 
Tolman-Oppenheimer-Volkoff (TOV) equations~\citep{tov1,tov2}. To solve such equations, we take the 
$\beta$-equilibrated energy density and pressure under chemical equilibrium and charge neutrality 
given by the \mbox{DD-vdW} model, along with the Baym-Pethick-Sutherland (BPS) equation of 
state~\citep{glend,bps} in the low density regime, namely, at $0.1581\times 
10^{-10}$~fm$^{-3}<\rho<0.008 907$~fm$^{-3}$, in order to specifically describe the NS crust. 
Certainly, a more profound study regarding a detailed description of the crust (nonuniform and 
clustered matter) of the neutron star~\citep{rmp,cc1,cc2,cc3,cc4}, even considering the pasta 
structure in the inner crust, is important and addressed to future works. Here we used a simple 
equation of state, the BPS one, connected to the \mbox{DD-vdW} model as a first approach to describe 
the stellar matter. In Fig.~\ref{mr}, we show the mass-radius diagrams obtained from the 
\mbox{DD-vdW} model. 

From the figure, one can see that the \mbox{DD-vdW} parametrizations are consistent with the 
findings related to the \mbox{PSR J0348+0432} pulsar, namely, $M_{\mbox{\tiny 
NS}}=(2.01\pm0.04)M_\odot$~\citep{psr2}, and also with the calculations performed by Steiner 
\citep{q01} and N\"attili\"a \citep{nat}. Another observational range for the NS mass is given in 
Ref.~\citep{psr1} and is related to the PSR J1614-2230 pulsar, in which $M_{\mbox{\tiny 
NS}}=(1.97\pm0.04)M_\odot$. These values were recently modified (shifted down) to $M_{\mbox{\tiny 
NS}}=(1.928\pm0.017)M_\odot$ in Ref.~\citep{fonseca}. The range of $K_0$ for the parametrizations 
compatible with the observational constraint of $M_{\mbox{\tiny NS}} \sim 2M_\odot$ is 
$247.8~\mbox{MeV}\leqslant K_0 \leqslant 251.5~\mbox{MeV}$, with the symmetry energy at the 
saturation density limited to $25~\mbox{MeV}\leqslant J \leqslant 35~\mbox{MeV}$. Once again, we 
restricted the curves to a density range compatible with the causal limit. For example, for the 
parametrization in which $J=25$~MeV and $K_0=247.8$~MeV, the causal limit allows the density range 
up to $\rho_/\rho_0= 5.66$. On the other hand, the maximum density reaches a value of $\rho_/\rho_0 
= 5.00$ for the parametrization in which $J=35$~MeV and $K_0=251.5$~MeV. In the figure, we also 
display the predictions of the LIGO/Virgo Collaboration for the radii related to the masses of the 
binary neutron star system of the GW170817 event. We see that the \mbox{DD-vdW} parametrizations are 
also consistent with this constraint.

\subsection{Tidal deformability calculations}

In the stellar matter context, the tidal deformability (TD) is the measure of the 
deformation in an NS due to an external field. In case of a binary NS system, the TD in 
one star is due to the gravitational field created by its companion. In a brief 
mathematical language, one can say that $Q_{ij} = -\lambda{\varepsilon}_{ij}$ is the 
relationship between the TD ($\lambda$), the quadrupole moment $Q_{ij}$ developed in the 
NS, and the external tidal field 
${\varepsilon}_{ij}$~\citep{tani08,tani082,tani10,damour}. The dimensionless TD 
($\Lambda$) is written in terms of the Love number $k_2$ as $\Lambda=2k_2/(3C^5)$, where 
$C=M_{\mbox{\tiny NS}}/R$ is the compactness of the NS and $R$ is its radius. In terms of 
$y_R\equiv y(R)$, the Love number is given by
\begin{eqnarray}
k_2 &=&\frac{8C^5}{5}(1-2C)^2[2+2C(y_R-1)-y_R]\times \nonumber\\
&\times&\Big\{2C [6-3y_R+3C(5y_R-8)] \nonumber\\
&+& 4C^3[13-11y_R+C(3y_R-2) + 2C^2(1+y_R)]\nonumber\\
&+& 3(1-2C)^2[2-y_R+2C(y_R-1)]{\rm ln}(1-2C)\Big\}^{-1}
\end{eqnarray}
where $y(r)$ is found through the solution of $r(dy/dr) + y^2 + y F(r) + r^2Q(r) = 
0$ simultaneously with the TOV equations, with $F(r) = \{1 - 4\pi r^2[\epsilon(r) - 
p(r)]\}/f(r)$,
\begin{eqnarray}
Q(r)&=&\frac{4\pi}{f(r)}\left[5\epsilon(r) + 9p(r) + 
\frac{\epsilon(r)+p(r)}{v_s^2(r)}- \frac{6}{4\pi r^2}\right]
\nonumber\\ 
&-& 4\left[ \frac{m(r)+4\pi r^3 p(r)}{r^2f(r)} \right]^2,
\end{eqnarray}
and $f(r)=1-2m(r)/r$. The inputs for $\epsilon(r)$ and $p(r)$ are given by 
Eqs.~(\ref{det})-(\ref{presst}), and the resulting neutron star mass for each radius 
$R$ is $M_{\mbox{\tiny NS}}=m(R)$.

\begin{figure}[!htb]
\centering
\includegraphics[scale=0.33]{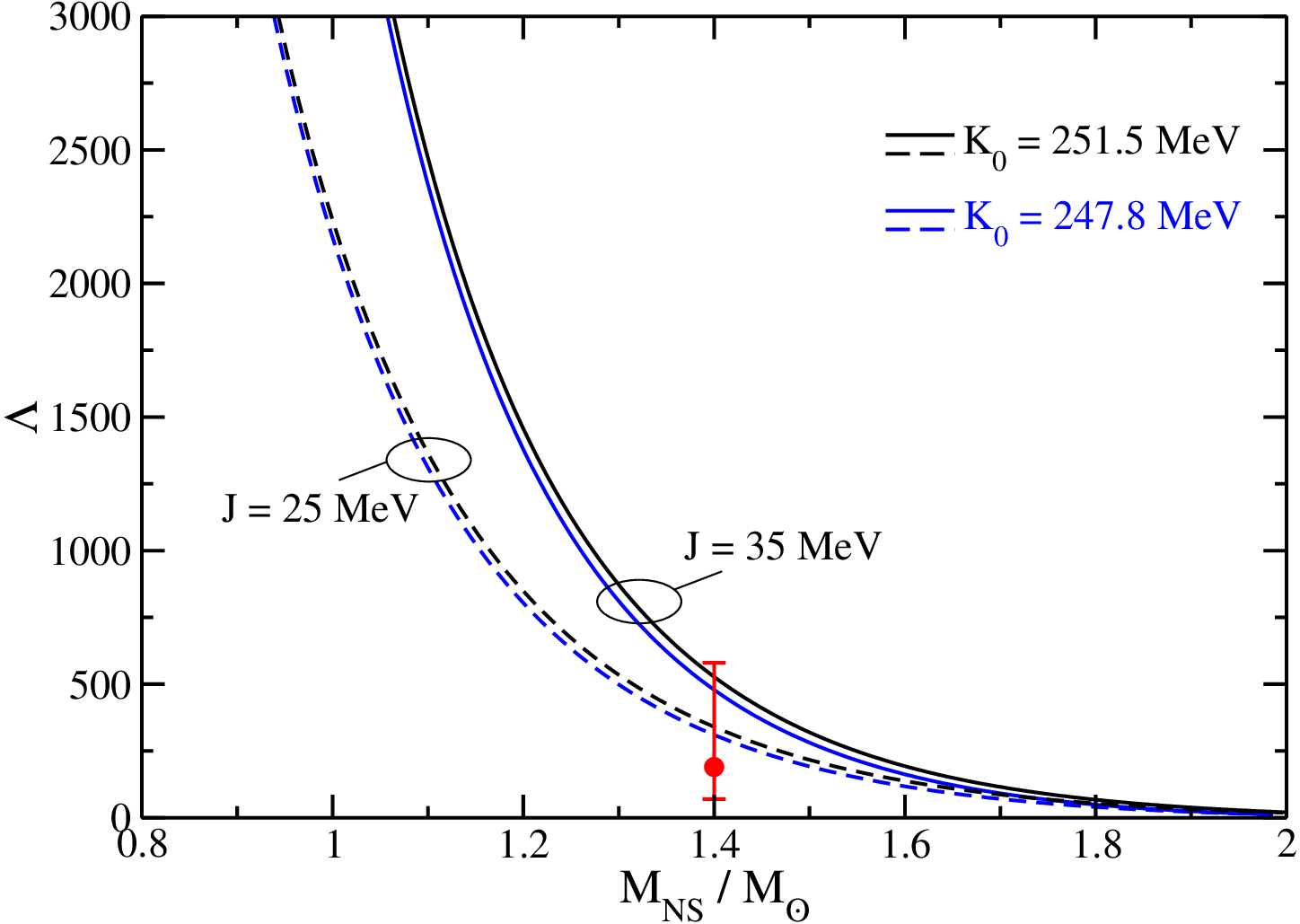}
\vspace{-0.2cm}
\caption{$\Lambda$ as a function of $M_{\mbox{\tiny NS}}$ for \mbox{DD-vdW}  
parametrizations that produce massive neutron stars. Red circle with error bar: 
constraint predicted in Ref.~\citep{ligonew}.}
\label{lambdamass}
\end{figure}

In Fig.~\ref{lambdamass} we display the dimensionless TD of the single NS as a function 
of its mass, for those \mbox{DD-vdW} parametrizations predicting massive NS as found in 
Fig.~\ref{mr}. From the figure, one can notice that $\Lambda$ decreases nonlinearly as the 
NS mass increases. Furthermore, the TD of a canonical neutron star, $\Lambda_{1.4}$, can 
also provide a constraint on the EoS. The recent LIGO/Virgo detection of gravitational 
waves suggests values for $\Lambda_{1.4}$ inside the range of 
$70\leqslant\Lambda_{1.4}\leqslant580$~\citep{ligonew}, which help us to test 
\mbox{DD-vdW} model. Our findings show that all parametrizations presented in 
Fig.~\ref{mr} predict $\Lambda_{1.4}$ inside this range. Quantitatively, the maximum value 
of this quantity for the \mbox{DD-vdW} parametrizations analyzed is 
$\Lambda_{1.4}=527$ (for $K_0=251.5$~MeV and $J=35$~MeV), completely inside 
the limits from the LIGO/Virgo collaboration. We can also verify from 
Fig.~\ref{lambdamass} that $\Lambda_{1.4}$ increases as both~$K_0$ or~$J$ increase.  
\begin{figure}[!htb]
\centering
\includegraphics[scale=0.33]{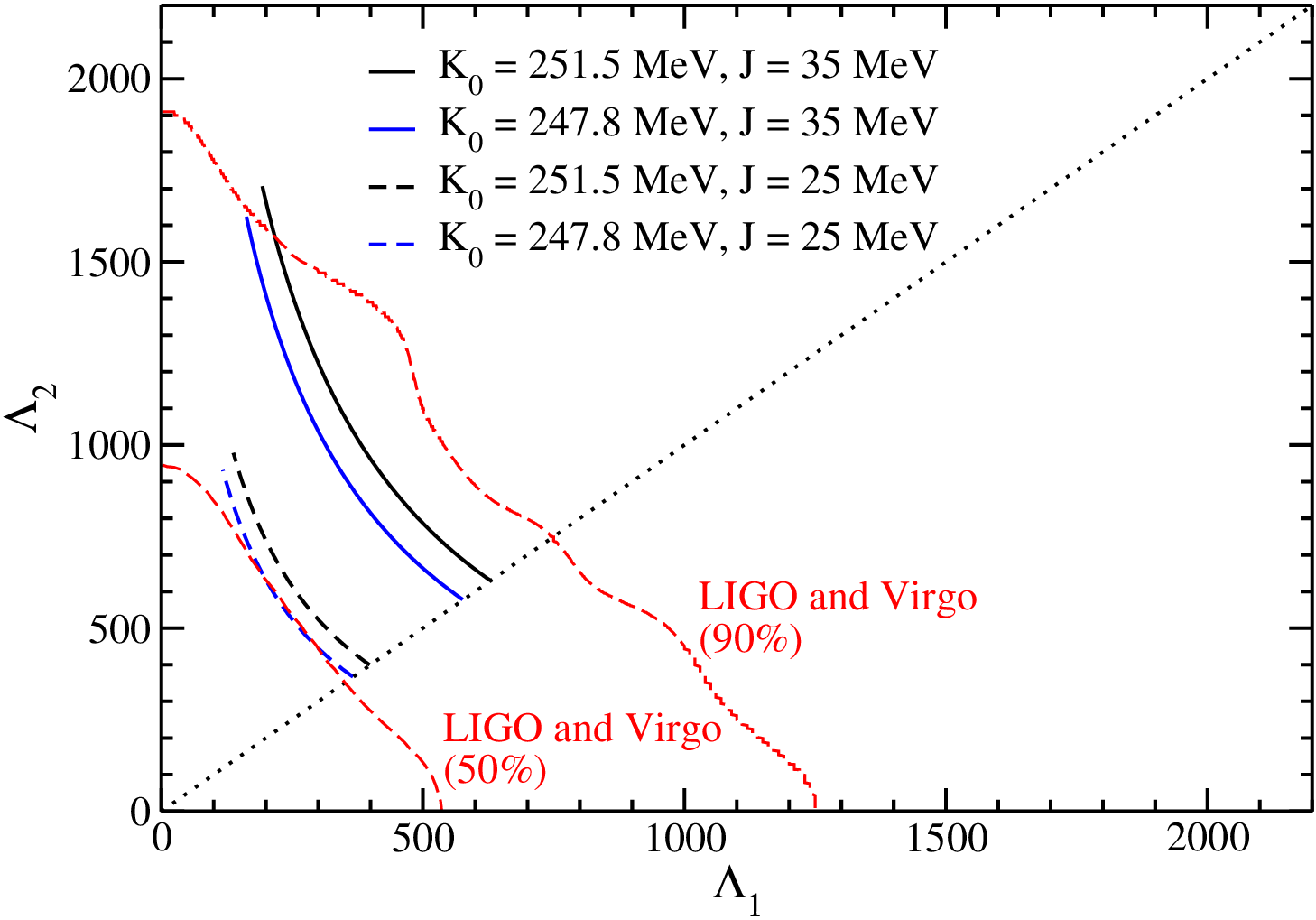}
\vspace{-0.2cm}
\caption{Tidal deformabilities for a binary NS system predicted by the \mbox{DD-vdW} 
model. Red dashed lines: 90\% and 50\% confidence lines related to the GW170817 event 
reported by the LIGO/Virgo collaboration~\citep{ligonew}.}
\label{lambda12}
\end{figure}

In Fig.~\ref{lambda12}, we show the dimensionless tidal deformabilities $\Lambda_1$ 
and $\Lambda_2$ for a binary NS system having primary mass $m_1$ and secondary mass 
$m_2$, respectively. At low orbital and gravitational frequency, the time evolution of the 
latter is determined by a defined combination given by ${\cal M} = (m_1 
m_2)^{3/5}/(m_1+m_2)^{1/5}$, with ${\cal M}$ being the chirp mass. In this work we fixed 
the chirp mass at ${\cal M}=1.188 M_{\odot}$ and run $m_1$ in the range of $1.36\leqslant 
m_1/M_{\odot} \leqslant 1.60$ (corresponding to $1.17\leqslant m_2/M_{\odot} \leqslant 
1.36$) according to the data from Ref.~\citep{ligonew2,ligonew}. In the figure, we also 
present two lines corresponding to the 90\% and 50\% confidence limits obtained from the 
LIGO/Virgo collaboration coming from the analysis of the GW170817 event~\citep{ligonew}. 
It is interesting to note that the tidal deformabilities for all analyzed \mbox{DD-vdW} 
parametrizations are inside the 90\% credible region. This interesting result incentive us 
to perform further studies concerning our new developed \mbox{DD-vdW} model also for other 
hadronic environments. 

\section{Summary and Concluding Remarks}
\label{con}

We developed a density-dependent van der Waals (\mbox{DD-vdW}) model by adopting the 
Carnahan-Starling method of excluded volume over the original vdW model. We also proposed 
a specific density dependence for the attractive part of the interaction by adding the $n$ 
power shown in Eq.~(\ref{arhodd}). Such a new parameter was included as a generalization 
of the Clausius model proposed in Ref.~\citep{vov3} to weaken the attraction and make our 
proposed model get closer to the Fermi free gas. We have shown in Figs.~\ref{vs2dd} 
and~\ref{flow} that such a weakening is important in both cases, namely, to push the break 
of the causal limit to higher densities, and to impose agreement with the high-density 
behavior of the thermodynamical pressure established in Ref.~\citep{danielewicz} (flow 
constraint). This constraint is significant and widely used to built or even to select 
relativistic hadronic models~\citep{rmf,new}. Furthermore, the incompressibility at the 
saturation density, $K_0$, is also controlled by fixing $n$. In our model we can ensure 
values for $K_0$ in the range of $220\,\mbox{MeV}\leqslant K_0 \leqslant 
260~\mbox{MeV}$, according to the current consensus for the value of this quantity~\citep{k4}. 

We also performed a generalization in the model in order to describe asymmetric systems.  
It was done by introducing a new term in the energy density with one more free parameter, 
adjusted to reproduce the symmetry energy at the saturation density. This new term can be 
seen as simulating the $\rho$ meson exchange between the finite nucleons (excluded volume 
included). Even with such a simple assumption, the model was shown to be compatible with 
some asymmetric nuclear matter constraints. Indeed, such a coupling can be improved, even 
with the inclusion of more free parameters, but we consider it quite suitable for a first 
approach.

Equations of state for neutron star matter under charge neutrality and 
$\beta$-equilibrium condition were also calculated with the \mbox{DD-vdW} model. It was 
shown that the mass-radius diagrams are in good agreement with the x-ray observations 
performed by Steiner~\citep{q01} and N\"attil\"a~\citep{nat}. The model is also compatible 
with the predictions for the binary neutron star system concerning the mass-radius 
diagrams obtained by the LIGO/Virgo collaboration~\citep{ligonew}. We also verified that 
the model predicts massive stars in the range of $(1.97-2.07)M_{\odot}$, in 
agreement with observational data from Ref.~\citep{psr2}. Furthermore, we calculated 
the dimensionless tidal deformability of a single neutron star as a function of its mass 
and found a maximum value of $\Lambda_{1.4}=527$ for the canonical star. 
This value and other ones also obtained from the model are entirely consistent with the 
range of $70\leqslant\Lambda_{1.4}\leqslant580$ recently obtained by the LIGO/Virgo 
collaboration~\citep{ligonew}. Moreover, we have calculated the tidal deformabilities 
$\Lambda_1$ and $\Lambda_2$ related to a binary system and verified that the results 
obtained are fully compatible with the observational boundaries established from the 
analysis of the neutron star merger event GW170817~\citep{ligonew}.

In summary, the results obtained from the calculations performed with our new proposed 
\mbox{DD-vdW} model are in agreement with the available predictions for symmetric,  
asymmetric and stellar matter, including data from the recent GW170817 event.

\section*{Acknowledgments} 

This work is a part of the project INCT-FNA Proc. No. 464898/2014-5, partially supported by Conselho 
Nacional de Desenvolvimento Científico e Tecnológico (CNPq) under grants 310242/2017-7 and 
406958/2018-1 (O. L.) and 433369/2018-3 (M.D.), by FAPESP under thematic projects 2014/26195-5 (M. 
B.), 2013/26258-4 (O. L.), and 2017/05660-0 (M. B., M. D., O. L.), and by the National key R\&D 
Program of China under grant 2018YFA0404402. O. L. and M. D also thank J. Piekarewicz for fruitful 
discussions and useful suggestions. O. L. , M. D. and B. M. S. dedicate this paper to Antonio 
Delfino Jr., from Federal Fluminense University (UFF), Brazil, who passed away in July 2019.


\begin{thebibliography}{}

%%%%%%%%%%%% Check arvix's
%%%%%%%%%%%%%%%%%%%%%%%%%%%%%%%%%%%%%%%%%%%%%%%
%%%%%%%%%%%%%%%%%%%%%%%%%%%%%%%%%%%%%%%%%%%%%%%

\bibitem[Abbot et al. (2017)]{ligonew2} 
Abbott, B.~P., et al. (LIGO Scientific Collaboration \& Virgo Collaboration) \ 2017, \prl, 119, 161101

\bibitem[Abbot et al. (2018)]{ligonew} 
Abbott, B.~P., et al. (LIGO Scientific Collaboration \& Virgo Collaboration) \ 2018, \prl, 121, 161101

\bibitem[Agrawal et al. (2005)]{agra05} Agrawal, B. K., Shlomo, S., \& Au, V. K.\ 2005, 
\prc, 72, 014310 

\bibitem[Ambartsumyan et al. (1960)]{amba60} Ambartsumyan, V.~A., \& Saakyan, G.~S.\ 
1960, \sovast, 4 187

\bibitem[Antoniadis et al. (2013)]{psr2} Antoniadis, J., Freire, P. C. C., Wex, N., et 
al.\ 2013, Sci 340, 1233232 

\bibitem[Atta \& Basu (2014)]{cc2} Atta, D., \& Basu, D. N.\ 2014, \prc, 90, 035802

\bibitem[Baldo \& Burgio(2016)]{baldo} Baldo, M., \& Burgio, G. F.\ 2016, PrPNP, 91, 
203 

\bibitem[Baym et al. (1971)]{bps} Baym, G., Pethick, C., \& Sutherland, P.\ 
1971, \apj, 170, 299

\bibitem[Bhuyan (2015)]{bhu15} Bhuyan, M.\ 2015 \prc, 92, 034323

\bibitem[Bhuyan et al. (2018)]{bhu18} Bhuyan, M., Carlson, B. V., Patra, S. K., et al.\ 
2018, 
\prc, 97, 024322

\bibitem[Boguta \& Bodmer (1977)]{bogu77} Boguta, J., \& Bodmer, A.\ 1977, \nphysa, 292, 
413

\bibitem[Bugaev (2008)]{causality} Bugaev, K. A.\ 2008, \nphysa, 807, 251

\bibitem[Bugaev et al. (2019)]{Bugaev:2018uum} Bugaev, K. A., Ivanytskyi, A. I., Sagun, V. V., 
Grinyuk, B. E., et al.\ 2019, Universe, 5, 63 


\bibitem[Carnahan \& Starling (1969)]{cs} Carnahan, N. F., \& Starling, K. E.\ 1969, 
\jcp, 51, 635

\bibitem[Carreau et al. (2019)]{cc4} Carreau, T., Gulminelli, F., \& Margueron, J.\ 2019,
arXiv:1902.07032

\bibitem[Chen et al. (2009)]{k2} Chen, L. W., Cai, B.-J., Ko, C. M, Li, B.-A., Shen, C., \& Xu, J.\ 
2009, \prc, 80, 014322

\bibitem[Colo et al. (2004)]{colo04} Colo, G., Giai, N. V., Meyer, J., et al.\ 2004 \prc, 
70, 024307 

\bibitem[Collins \& Perry (1975)]{coll75} Collins, J. C., \& Perry, M. J.\ 1975, \prl, 
34, 1353 

\bibitem[Damour et al. (2012)]{damour} Damour, T., Nagar, A., \& Villain, L.\ 2012, 
\prd, 85, 123007

% \bibitem[Danielewicz (2003)]{dani03}\sout{Danielewicz, P.\ 2003, \nphysa, 727, 
% 233}

\bibitem[Danielewicz et al. (2002)]{danielewicz} Danielewicz, P., Lacey, R., \& Lynch, W. 
G.\ 2002, Sci, 298, 1592

\bibitem[Demorest et al. (2010)]{psr1} Demorest, P. B., Pennucci, T., Ransom, S. M., et 
al.\ 2010, Natur, 467, 1081

\bibitem[Ducoin et al. (2011)]{cc1} Ducoin, C., Margueron, J., Provid\^encia, C., et al.\ 
2011, \prc, 83, 045810

\bibitem[Dutra et al. (2014)]{rmf} Dutra, M., Louren\c{c}o, O., Avancini, S. S., et al.\ 
2014, \prc, 90, 055203

\bibitem[Dutra et al. (2012)]{skyrme} Dutra, M., Louren\c{c}o, O., Martins, J. S. 
S\'a, et al.\ 2012 \prc, 85, 035201

\bibitem[Flanagan \& Hinderer (2008)]{tani082} Flanagan, E. E., \& Hinderer, T.\ 2008, 
\prd, 77, 021502

\bibitem[Fonseca et al. (2016)]{fonseca} Fonseca, E., Pennucci, T. T., Ellis, J. A., et al.\ 2016, 
\apj, 832, 167

\bibitem[Garg \& Colo (2018)]{k4} Garg, U., \& Colo, G.\ 2018, PrPNP, 101, 55

\bibitem[Glendenning (1985)]{glen85} Glendenning, N. K.\ 1985, \apj, 293, 470

\bibitem[Glendenning (2000)]{glend} Glendenning, N.~K.\ 2000, Compact Stars (Springer:New 
York)

\bibitem[Glendenning \& Moszkowski (1991)]{glen91} Glendenning, N. K., 
\&  Moszkowski, S. A.\ 1991, \prl, 67, 2414 

\bibitem[Glendenning \& Schaffner-Bielich (1998)]{glen98} Glendenning, N. K., \& 
Schaffner-Bielich, J.\ 1998, \prl, 81, 4564

\bibitem[Glendenning \& Schaffner-Bielich (1999)]{glen99} Glendenning, N. K., \&  
Schaffner-Bielich, J.\ 1999, \prc, 60, 025803

\bibitem[Gupta \& Arumugam (2012)]{gupt12} Gupta, N., \&  Arumugam, P.\ 2012, \prc, 
85, 015804

\bibitem[Hinderer (2008)]{tani08} Hinderer, T.\ 2008, \apj, 677, 1216 
 
\bibitem[Hinderer et al. (2010)]{tani10} Hinderer, T., Lackey, B. D., Lang, R. N., et 
al.\ 2010, \prd, 81, 123016

\bibitem[Holt \& Lim (2018)]{lim} Holt, J. W., \& Lim, Y.\ 2018, PhLB, 784, 77

\bibitem[Horowitz et al. (2014)]{se6} Horowitz, C. J., Brown, E.F., Kim, Y., Lynch, W. G., Michaels, 
R., Ono, A., Piekarewicz, J., Tsang, M. B., \& Wolter,  H. H.\ 2014, JPhG, 41, 093001

\bibitem[Horowitz \& Piekarewicz (2001)]{pieka} Horowitz C. J., \& Piekarewicz, J.\ 2001, 
\prl, 86, 5647 

% \bibitem[Jiang et al. (2012)]{jian12}\sout{Jiang, H., Fu, G.J., Zhao, Y.M., 
% et al.\ 2012, \prc, 85, 024301 }

\bibitem[Kumar et al. (2018)]{new} Kumar, B., Patra, S. K., \& Agrawal, B. K.\ 2018, 
\prc, 97, 045806

\bibitem[Lattimer \& Prakash (2004)]{latt04} Lattimer, J. M., \& Prakash, M.\ 2004, 
Sci 304, 536
 
\bibitem[Lalazissis et al. (2009)]{nl3s} Lalazissis, G. A., Karatzikos, S., Fossion, R. 
et al.\ 2009, PhLB, 671, 36

\bibitem[Li (2017)]{se5} Li, B. A.\ 2017, NPN, 27, No. 4, 7

\bibitem[Li et al. (2008)]{se1} Li, B. A., Chen, L. W., \& Ko, C. M.\ 2008 \physrep, 464, 113

\bibitem[Li \& Han (2013)]{jl} Li, B.-A., \& Han, X.\ 2013, PhLB, 727, 276

\bibitem[Li et al. (2019)]{se7} Li, B.-A., Krastev, P. G., Wen, D.-H., \& Zhang, N.-B.\ 2019, EPJA, 
55, 117

\bibitem[Li et al. (2014)]{se4} Li, B.A., Ramos, \`A., Verde, G., \& Vida\~na, I.\ 2014, EPJA, 50, 
2

\bibitem[Lynch et al. (2009)]{se2} Lynch, W. G., Tsang, M. B., Zhang, Y., Danielewicz, P., Famiano, 
M., Li, Z., \& Steiner, A. W.\ 2009, PrPNP, 62, 427

\bibitem[Louren\c{c}o et al (2007)]{ijmpe1} Louren\c{c}o, O., Dutra, M., Delfino, A., et 
al.\ 2007, IJMPE, 16, 3037

\bibitem[Louren\c{c}o et al. (2016)]{ijmpe2} Louren\c{c}o, O., Santos, B. M., Dutra, 
M.\ 2016 \prc, 94, 045207

 
\bibitem[N\"attili\"a et al. (2016)]{nat} N\"attili\"a, J., Steiner, A. W., Kajava, J. J. 
E., et al.\ 2016, \aap, 591, A25 

\bibitem[Oertel et al. (2017)]{rmp} Oertel, M., Hempel, M., Kl\"ahn, T., et al.\ 2017, 
RvMP, 89, 015007

\bibitem[Oppenheimer \& Volkoff (1939)]{tov2} Oppenheimer, R., \& Volkoff, G. M.\ 1939 
PhRv, 55, 374

\bibitem[Pearson et al. (2010)]{k3} Pearson, J. M., Chamel, N., \&  Goriely, S.\ 2010, \prc, 82, 
037301

\bibitem[Rueda et al. (2014)]{cc3} Rueda, J. A., Ruffini, R., Wu, Y.-B., et al.\ 2014, 
\prc, 89, 035804

\bibitem[Sagun et al. (2018)]{Sagun:2017eye} Sagun, V. V., Bugaev, K. A., Ivanytskyi, A. I., 
Yakimenko, I. P., et al.\ 2018, EPJA, 54, 100

\bibitem[Santos et al. (2014)]{bianca} Santos, B. M., Dutra, M., Louren\c co, O., \& 
Delfino, A.\ 2014, \prc, 90, 035203

\bibitem[Santos et al. (2015)]{bianca2} Santos, B. M., Dutra, M., Louren\c{c}o, O., \& 
Delfino, A.\ 2015, \prc, 92, 015210

\bibitem[Serot \& Walecka(1979)]{sero79} Serot, B. D., \& Walecka, J. D.\ 1979, PhLB, 
87, 172.

\bibitem[Shlomo \& Youngblood (1993)]{k1} Shlomo, S., \& Youngblood, D. H\ 1993, \prc, 47, 529

\bibitem[Skyrme (1959)]{skyr59} Skyrme, T. H. R.\ 1959, NuPh, 9, 615

\bibitem[Steiner et al. (2010)]{q01} Steiner, A. W., Lattimer, J. M., \& Brown, E. F.\ 
2010, \apj, 722, 331

\bibitem[Steiner et al. (2005)]{se3} Steiner, A. W., Prakash, M., Lattimer, J. M., \& Ellis, P. J.\ 
2005, \physrep, 411, 325

\bibitem[Stone et al. (2014)]{stone} Stone, J. R., Stone,  N. J., \&  Moszkowski S. 
A.\ 2014, \prc, 89, 044316

\bibitem[Todd-Rutel \& Piekarewicz (2005)]{todd05} Todd-Rutel, B. G., \& 
Piekarewicz, J.\ 2005, \prl, 95, 122501 

\bibitem[Tolman (1939)]{tov1} Tolman, R. C.\ 1939, PhRv, 55, 364

\bibitem[Vautherin \&  Brink (1972)]{vaut72} Vautherin, D., \&  Brink, D. M.\ 1972, \prc, 5, 626

\bibitem[Vovchenko (2017)]{vov3} Vovchenko, V.\ 2017, \prc, 96, 015206 

\bibitem[Vovchenko et al. (2015a)]{vov1} Vovchenko, V., Anchishkin, D. V., \& Gorenstein, 
M. I.\ 2015a, \prc, 91, 064314
 
\bibitem[Vovchenko et al. (2015b)]{vov2} Vovchenko, V., Anchishkin, D. V., \& Gorenstein, 
M. I.\ 2015b, JPhysA, 48, 305001
 
\bibitem[Vovchenko et al. (2018)]{vov4} Vovchenko, V., Gorenstein, M. I., \& Stoecker, 
H.\ 2018, EPJA, 54, 16 

\bibitem[Vovchenko et al. (2017)]{multi} Vovchenko, V., Motornenko, A., Alba, P., et 
al.\ 2017, \prc, 96, 045202

\bibitem[Walecka (1974)]{wale74} Walecka, J.\ 1974, AnPhy, 83, 491

\end{thebibliography}
\end{document}